\documentclass{PoS}
\usepackage{multicol}
\title{Recent Progress on the QCD Phase Diagram}

\ShortTitle{Recent Progress on the QCD Phase Diagram}

\author{Sayantan Sharma\\
        The Institute of Mathematical Sciences, Chennai 600113, India\\
        E-mail: \email{sayantans@imsc.res.in}}

\abstract{Recent progress and the latest results on the bulk thermodynamic properties of QCD matter from lattice is
reviewed.
In particular I will stress upon the fact that lattice techniques are now entering into precision era where they can provide us with new 
insights on even the microscopic degrees of freedom in different phases of QCD. I will discuss some instances from the recent 
studies of topological fluctuations and screening masses. The progress towards understanding the effects of anomalous $U_A(1)$ 
symmetry on the chiral crossover transition and transport properties of QCD matter will also be discussed.}

\FullConference{The 36th Annual International Symposium on Lattice Field Theory - LATTICE2018\\
		22-28 July, 2018\\
		Michigan State University, East Lansing, Michigan, USA.}

\begin{document}

\section{Introduction}
The QCD phase diagram has driven the scientific curiosity of the community for more than thirty years, from 
its earliest versions discussed as early as 1980s. Understanding the phase diagram allows us to explain the 
origin of mass of 99.9\% of the visible matter in the present universe. It has motivated large scale experiments 
from the LHC to the RHIC at BNL and is the special focus of the BES II runs at the RHIC during 2019-20. Several 
upcoming experiments at FAIR, NICA and JPARC are being designed to probe the phase diagram at very high baryon 
densities, yet to be understood.
Experimental challenges aside, it is one of the most challenging problems in theoretical physics. Lattice studies have 
produced some of the remarkable results till now; it has now conclusively demonstrated that the phase transition at vanishingly 
small baryon densities is a smooth crossover~\cite{Aoki:2006we,Bazavov:2011nk,Bhattacharya:2014ara}. Continuum results 
for bulk thermodynamic quantities like entropy density, pressure and the Equation of state (EoS) at zero baryon density 
are now known to very high precision~\cite{Borsanyi:2013bia} with new results on continuum 
estimates for the EoS available at baryon densities as large as $\mu_B/T\sim 2.5$~\cite{Bazavov:2017dus}.
Efforts are underway to develop new lattice techniques to extend these calculations to even larger baryon densities 
$\mu_B/T\sim 3$. Moving ahead with these successes, I will show some instances, how lattice techniques are 
becoming mature enough to extend beyond bulk thermodynamic observables to understand the more microscopic details of the 
different phases of QCD, in particular, the microscopic origins of chiral symmetry breaking and deconfinement.

The review is organized as follows: In the first section, the recent updates on the thermodynamics crossover transition 
at $\mu_B=0$ are discussed. The anomalous $U(1)$ part of the softly broken chiral symmetry in QCD is believed to play an 
important role in determining the nature of the chiral phase transition in the limit when up and down quark masses are vanishingly 
small~\cite{Pisarski:1983ms}. I will discuss the latest lattice results on the fate of $U_A(1)$ anomalous symmetry and 
how its origin of it can be traced back to the non-trivial topology of QCD. This leads to the next section which elaborates 
how the lattice community is trying to learn more about QCD phase diagram by varying the mass and the number of quark flavors 
within the so-called Columbia plot. The Columbia plot is now extensively studied along a new axis, by including an imaginary 
chemical potential to the QCD action. Finally I discuss how both imaginary chemical potential techniques as well as Taylor 
expansion in $\mu_B$ is allowing us to sketch the phase diagram in the finite density regime and possibly constrain 
a region in $T$-$\mu_B$ plane which may have the critical end-point. Other interesting topics discussed in the finite 
temperature sessions not included in this review are QCD at finite magnetic fields~\cite{tomiya}, strong coupling QCD~\cite{str} and QCD thermodynamics 
at large N~\cite{largeN}.

\section{Symmetries and phase diagram at $\mu_B=0$}
Since the up and the down quark masses are much lighter than the intrinsic scale of QCD i.e $ m_l=m_{u,d} << \Lambda_{QCD}$, the 
$U_L(2)\times U_R(2)$ symmetry of the action is very mildly broken. $U_L(2)\times U_R(2)$ is isomorphic to $SU(2)_V\times 
SU(2)_A\times U_B(1)\times U_A(1)$ and 2+1 flavor QCD, has to a very good approximation, a $SU(2)_V\times SU(2)_A\times U_B(1)$ 
symmetry which is broken to $SU(2)_V\times U_B(1)$ leading to chiral symmetry breaking. The anomalous $U_A(1)$ part is always broken 
due to quantum effects. Though the chiral symmetry is exact in the limit $m_u,m_d\rightarrow0$, however remnants of it exist in 
chiral observables. For example, it was discussed in this conference~\cite{plattice} that the temperature 
at which an inflection point exist for the subtracted chiral condensate is consistent with the one at which the chiral susceptibility or its disconnected part peaks. An unweighted average of all these temperatures have been calculated in the continuum limit which allows for a more precise determination of the pseudo-critical temperature $T_c=156.5\pm 1.5$ MeV~\cite{Steinbrecher:2018phh}. In contrast to the earlier reported value of $T_c=154(9)$ MeV~\cite{Bazavov:2011nk} in the continuum limit, the systematic errors have reduced significantly by more than $80\%$. For recent updates on the results of chiral observables and measurement of $T_c$ with twisted mass fermions 
see~\cite{Burger:2018fvb}.

The $U_A(1)$ part is an anomalous symmetry, there is no corresponding order parameter. From renormalization group studies 
of model quantum field theories with same symmetries as QCD, it has been observed that the order of phase transition 
for 2 flavor QCD depends on whether $U_A(1)$ breaking effects survive or gets effectively 
restored at $T_c$~\cite{Pisarski:1983ms}. Further studies with epsilon expansion~\cite{Pelissetto:2013hqa} and 
conformal bootstrap~\cite{Nakayama:2014sba} have revealed a possibility of a first order or even a second order 
phase transition of $U_L(2)\times U_R(2)/U_V(2)$ universality if the $U_A(1)$ is effectively restored near $T_c$ 
in contrast to an $O(4)$ second order transition if it remains broken. The magnitude of effective breaking of $U_A(1)$ 
can only be answered non-perturbatively and lattice techniques have immensely contributed towards a more systematic 
understanding of this issue. In order to quantify the effects of $U_A(1)$ at $T_c$, it was suggested quite sometime back to look at 
the degeneracy of the integrated two-point correlation functions of iso-triplet pseudo-scalar and scalar mesons~\cite{Shuryak:1993ee}.
The integrated correlation functions can be written in terms of the eigenvalues $\lambda$ and density $\rho(\lambda)$ of 
the QCD Dirac operator  as $\chi_\pi-\chi_\delta=\int d\lambda \frac{4 m_l^2\rho(\lambda)}{(\lambda^2+m_l^2)^2}$,
hence the properties of the eigenvalue spectrum as a function of temperature tells us about the fate of the $U_A(1)$. 
One way to trivially realize $U_A(1)$ restoration along with the chiral symmetry is to have $\rho(\lambda\rightarrow 0)=0$. On the 
other hand if the eigenvalue density has non-analyticities in its infra-red spectrum like $m_l^\alpha \delta(\lambda), \alpha \in [0,2)$ 
then $\chi_\pi-\chi_\delta$ is non-zero even in the chiral limit~\cite{Bazavov:2012qja}. Recent theoretical studies suggest it is important to look at higher order correlation functions in all these mesonic quantum number channels~\cite{Aoki:2012yj}. In the chiral limit, calculations show that $U_A(1)$ breaking effects are invisible in upto 6-point correlation functions in the scalar-pseudo-scalar channel if the eigenvalue density goes as 
$\rho(\lambda)\sim \lambda^3$~\cite{Aoki:2012yj}. The main issues on the study of $U_A(1)$ addressed and reported in this conference are,
\begin{itemize}
  \item If one studies the eigenvalue spectrum of QCD at the physical point how does it quantitatively change 
 as one goes towards the chiral limit. Are these spectra very different?
 \item Status of the finite volume and finite cut-off effects that crucially affects these studies.
\end{itemize}

\begin{figure}[]
\begin{center}
\includegraphics[scale=0.6]{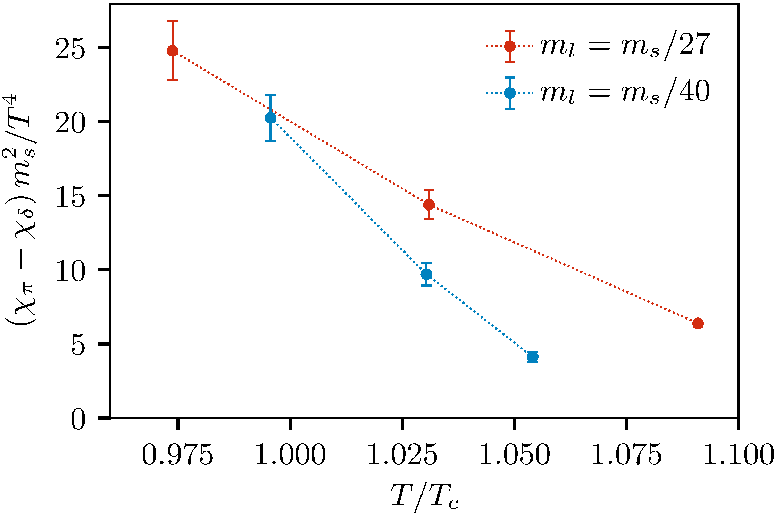}
\includegraphics[scale=0.25]{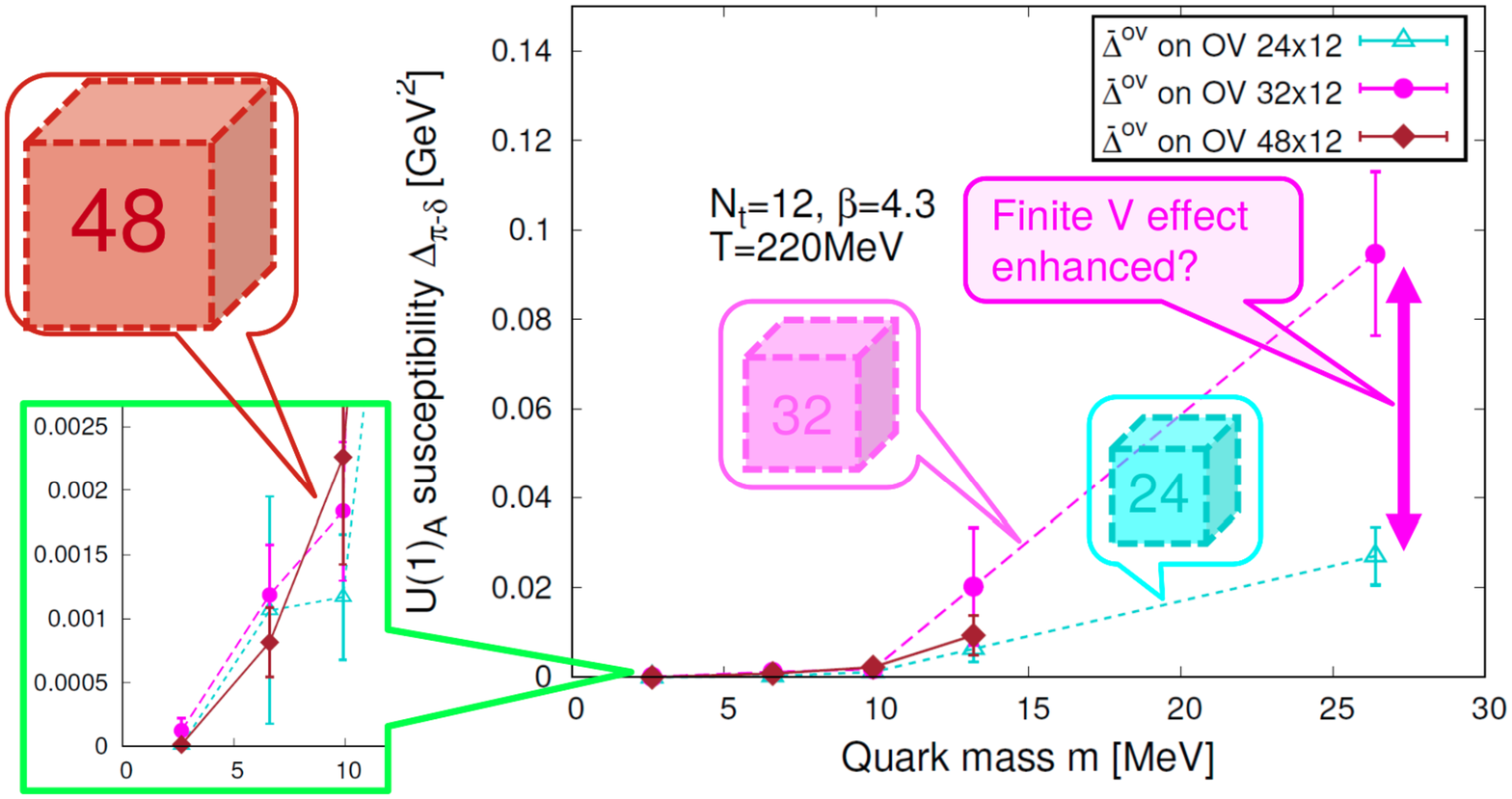}
\caption{The $U_A(1)$ breaking observable $m_s^2(\chi_\pi-\chi_\delta)/T^4$ as a function of $T$ for two different light quark 
masses from~\cite{lukaslat} (left panel). 
The status of the same observable as a function of quark mass and different volumes studied using reweighted M\"{o}bius domain wall 
fermions from ~\cite{suzukilatt} (right panel).}
\label{fig:chiral1}
\end{center}
\end{figure}

New results on $\chi_\pi-\chi_\delta$  in 2-flavor QCD were presented in this conference~\cite{suzukilatt}, 
summarized in right panel of Fig. \ref{fig:chiral1}. It is observed that as one approaches the chiral limit, the 
finite volume effects could be milder (right panel of Fig.  \ref{fig:chiral1}).
For the physical quark masses the $U_A(1)$ breaking is still finite on lattices of size $48^3\times 12$ which seems to 
decrease to zero in the limit $m_l\rightarrow 0$. It would be interesting to study in detail how this reweighting of domain wall 
configurations work at large volumes and towards the chiral limit. The other approach reported was to calculate the 
eigenvalues of QCD Dirac operator for 2+1 flavors by fixing the strange quark mass $m_s$ to its physical value and reducing the light 
quark masses towards the chiral limit. New results on the eigenvalue spectrum of overlap Dirac operator on gauge ensembles generated 
using Highly Improved Staggered Quark (HISQ) discretization reported in Ref.~\cite{lukaslat}, 
shows that the analytic part of the infrared spectrum is quite robust. $\rho(\lambda)\sim \lambda$ at around $1.1~T_c$ even when the 
light quark masses are reduced from $m_l=m_s/20$ \cite{Dick:2015twa} to $m_s/40$. A small non-analytic peak for $\lambda \rightarrow 0$ 
observed in the eigenvalue spectrum has been suspected due to effects of partial quenching~\cite{Tomiya:2016jwr}. In 
order of verify that the HISQ eigenvalue spectrum has been measured with the same valence and sea quark operators on fine lattices 
$64^3\times16$ just above $T_c$.  The small non-analytic peak seems to appear as one approaches the continuum limit~\cite{Sharma:2018syt}. 
It will be interesting to check this with other fermion discretizations like domain wall fermions also in the continuum limit 
though this study will be computationally much more intensive. As evident from the left panel of Fig. \ref{fig:chiral1} both analytic and 
non-analytic parts of $\rho(\lambda)$ contribute to $U_A(1)$ breaking ($\chi_\pi-\chi_\delta$ has been renormalized appropriately 
to ameliorate the effects of partial quenching) which seems to survive even for $m_l=m_s/40$ at temperatures upto $1.1~T_c$. 
This suggests $U_A(1)$ breaking survives towards the chiral limit~\cite{lukaslat}.

Another observable that measures the topological fluctuations of QCD vacuum is the topological susceptibility $\chi_t$. In LATTICE 2017, 
an extensive discussion of results from different groups suggest that for $T>3~T_c$ the temperature dependence of $\chi_t$ is consistent 
with the expectations from dilute instanton gas approximation (DIGA)~\cite{Petreczky:2016vrs,Frison:2016vuc,Borsanyi:2016ksw} whereas 
non-trivial temperature dependence is seen for $T_c <T< 3 ~T_c$~\cite{Bonati:2015vqz,Petreczky:2016vrs}. 
New results with twisted mass fermions in $2+1+1$ QCD also confirms this overall picture~\cite{Burger:2018fvb}. Though 
the temperature dependence of $\chi_t$ agrees quite well with DIGA for $T>3~T_c$, its magnitude  has to be scaled by a 
factor of $\sim 9$ to match with the leading order semi-classical result at $T\sim 450$ MeV~\cite{Borsanyi:2016ksw}. This 
is due to the fact that the semi-classical result includes the color screening function at LO which has a slow convergence 
with the coupling~\cite{Petreczky:2016vrs}. It was argued that the semi-classical expansion of instanton action may not be  
as uncontrolled at $T\gtrsim 1.5$ GeV~\cite{Dine:2017swf}. It would nevertheless be important to measure $\chi_t$ for $T>1$ GeV 
on the lattice to observe this convergence.
However it is assuring that in the context of axion mass estimation, the temperature dependence of $\chi_t$ plays the decisive 
factor~\cite{Petreczky:2016vrs}, changing the scale factor from $15$ to unity only changes the axion mass by $20\%$.

\begin{figure}[]
\begin{center}
\includegraphics[scale=0.18]{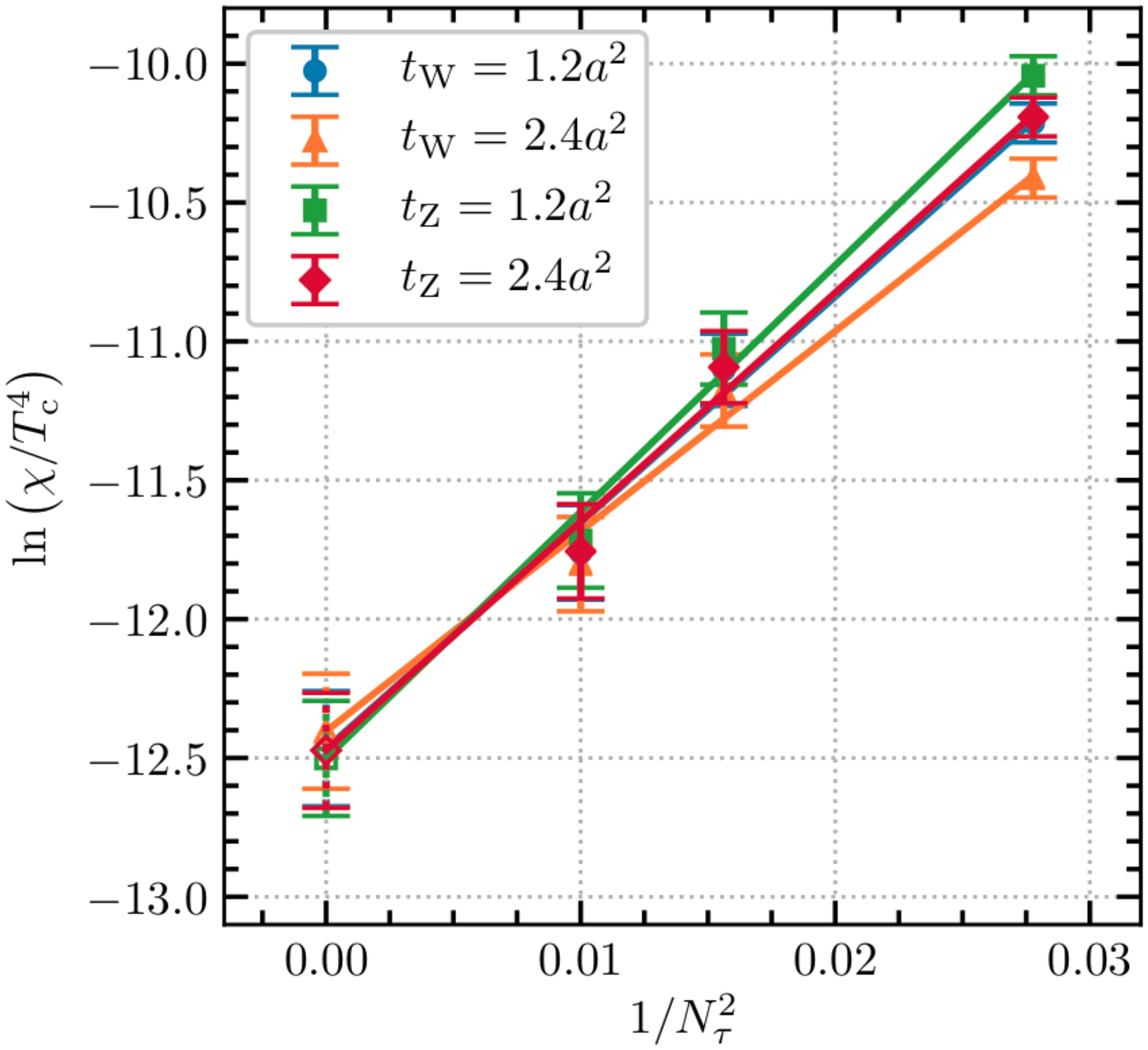}
\includegraphics[scale=0.17]{datiestrapolati.eps}
\includegraphics[scale=0.35]{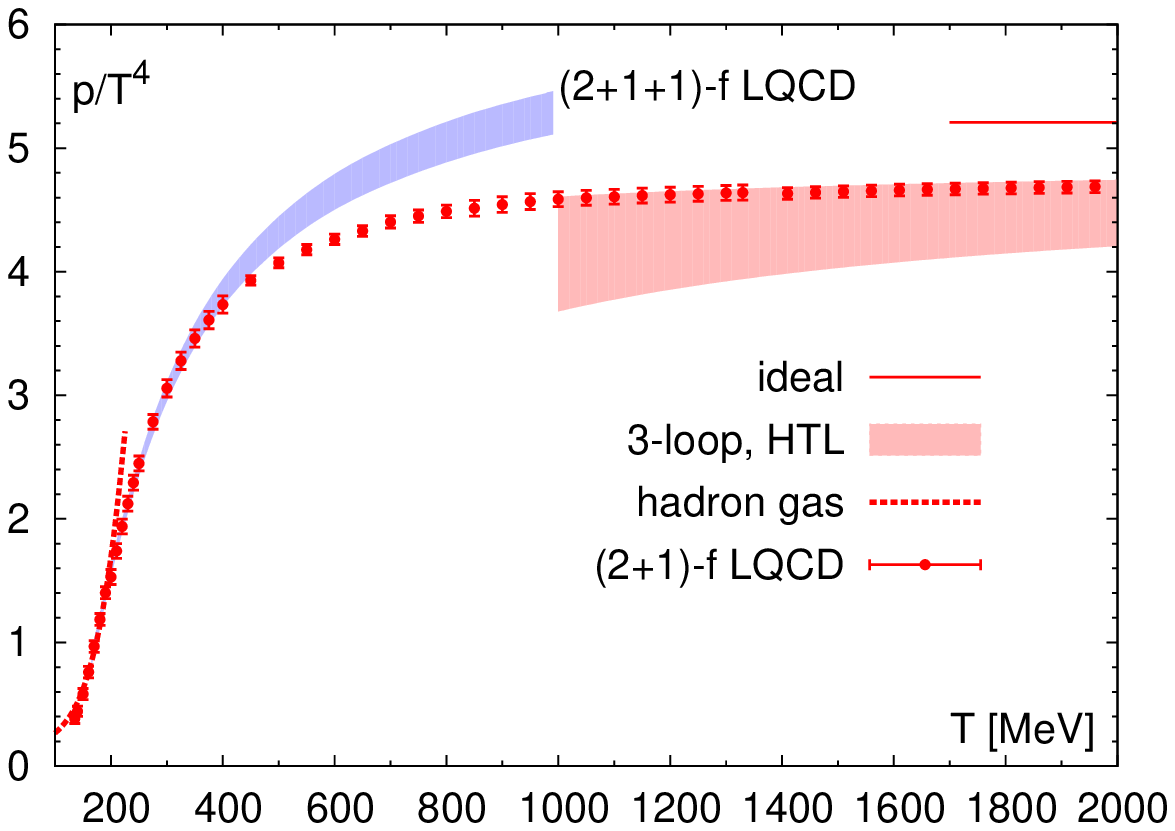}
\caption{Continuum extrapolated $\chi_t$ for $SU(3)$ gauge theory at $T\sim 4~T_d$ (left panel) from \cite{Jahn:2018dke}. The central 
panel contains continuum extrapolated value of $\chi_t$ in QCD with stout fermions by two independent analysis from \cite{Bonati:2018blm}. 
The right panel shows the QCD EoS upto $2$ GeV for $2+1$ flavor QCD from lattice with HISQ fermions from \cite{peter}, compared to EoS 
for $2+1+1$ QCD from \cite{Borsanyi:2016ksw} and HTL perturbative estimates.}
\label{fig:eos}
\end{center}
\end{figure}

Interesting algorithmic developments have been reported since LATTICE 2017 to measure $\chi_t$ to very high temperatures~\cite{Bonati:2017woi,Jahn:2018dke}. Since topological tunnelings become rarer as one goes to higher temperatures, one has to sample a large number of configurations to measure $\chi_t$ making the problem computationally challenging.  It has been shown that sampling the ensembles with a reweighting factor with coarse-grained definition of topological charge, reduces the probability to get stuck at one topological sector. New results on continuum extrapolated $\chi_t$ for pure gauge theory (in left panel of Fig. \ref{fig:eos}) at $\sim 4~T_d$ was reported~\cite{Jahn:2018dke}, which were calculated with very moderate computational efforts. Reweighting techniques have been applied to QCD with stout fermion discretization and $\chi_t$ has been calculated~\cite{Bonati:2018blm} after performing very careful finite volume and continuum extrapolation at $T\sim 450$ MeV, see central panel of Fig. \ref{fig:eos}. The results are consistent with the earlier results of $\chi_t$ calculated using a different reweighting 
technique~\cite{Frison:2016vuc} performed along the temperature axis starting from a low temperature ensemble~\cite{Borsanyi:2016ksw}. 
Whereas this finite temperature reweighting is expected to work well for pure gauge theory where the temperature dependence is along 
the expectations of DIGA beyond $T_c$, it is more non-trivial to extend this technique in full QCD where the $T$-dependence is more 
intricate than naive DIGA below $2.5 ~T_c$. It is assuring that new techniques~\cite{Bonati:2018blm} 
confirm earlier reported results. Several other algorithms in the context of quantum mechanics~\cite{Bonati:2017woi} are discussed 
which have potential to be applied to QCD, some techniques discussed earlier like metadynamics~\cite{Laio:2015era} requires more 
extensive application. 
The endeavor towards measuring rare topological fluctuations at high temperature QCD has motivated development of new lattice 
techniques which can be applied to a more general problem when one approaches the continuum limit, when irrespective of the temperature, 
the ensembles get stuck in one topological sector.

New studies on understanding the topology in QCD near the chiral crossover region have been reported in this conference. 
In fact higher moments of the free energy $F(\theta)$ are more sensitive to the microscopic topological objects~\cite{Bonati:2013tt}. 
It has been reported earlier that the fourth moment of $F(\theta)$ has a value that is different from DIGA in the 
range $T_c~<T~<2~ T_c$~\cite{Bonati:2015vqz}. This naturally leads to the question: what explains such an 
observation? At finite temperature the eigenvalues of the Polyakov loop at spatial infinity or the holonomy characterizes the 
properties of the instantons. For trivial holonomy the finite action solution at non-zero temperatures or calorons have been known for 
quite sometime~\cite{Harrington:1978ve}. Towards the end of 90's, the calorons with non-trivial holonomy were discovered~\cite{Kraan:1998sn}. 
In fact it was shown that calorons in $SU(N)$ gauge theory consists of $N$ dyons, which carry a fraction $1/N$ of the net topological 
charge. Additionally dyons carry both color electric and magnetic charges and they combine in a way that the calorons are charge neutral objects. 
Calorons with trivial holonomy cannot explain the mechanism of confinement in gauge theories, mean-field studies of dyon gas hints to the fact 
that they may be a key towards understanding confinement~\cite{Diakonov:2009jq}. It is therefore important not only to establish the existence 
of such objects non-perturbatively in QCD but understand their interactions. 
An new study has been reported in this conference~\cite{rlarsenlat} furthering the earlier studies on dyons~
\cite{Gattringer:2002wh,Bornyakov:2015xao}. QCD ensembles are generated during a Monte-Carlo evolution with (anti)-periodic 
boundary conditions imposed along the temporal direction for (fermions) gauge fields hence an isolated dyon cannot exist 
on the lattice. However zero-modes of the valence Dirac operator with a general boundary condition 
$\psi(\tau+\beta)=\rm{e}^{i\phi}\psi(\tau)$, such that the twist angle $\phi$ lies between the eigenvalues of Polyakov loop, 
will detect the dyon characterized by the difference between these eigenvalues. This technique has been used to detect 
and characterize the zero modes of the overlap operator with different boundary conditions on M\"{o}bius domain wall 
fermion sea ensembles at temperatures between $T_c<T<1.1~ T_c$~\cite{rlarsenlat}. In fact density profiles of the zero 
mode wavefunctions show a very good agreement with the analytic profiles of dyons and their characteristic fall-off at 
large distances have been suggested as one of the signatures to identify dyons on the lattice. From a detailed study of 
near-zero modes, the interactions between dyons have been inferred qualitatively~\cite{rlarsenlat}. 
These insights will eventually lead us to an understanding of the mechanism of deconfinement and the yet un-explained temperature 
variation of $\chi_t$ just above $T_c$. At higher temperatures, $T>2~T_c$, the holonomy is trivial but there may be localized 
fluctuations of the Polyakov loop value, which is conjectured to provide the 'disordered' landscape to localize the 
bulk eigenfunctions of the QCD Dirac operator~\cite{Bruckmann:2011cc}. A new study of the localization properties of the 
overlap Dirac operator on $2+1+1$ twisted mass sea ensembles (with pion mass of $\sim 370$ MeV) has been reported in this 
conference~\cite{Holicki:2018sms}. It further provides support for the conjecture that local negative fluctuations of the 
Polyakov loop provides the disorder required for localization of bulk eigenmodes and also reports that a dilute instanton 
gas cannot support such a localization~\cite{Holicki:2018sms}. 

Updates on the quark mass and volume dependence of $\chi_t$ in 2 flavor QCD with overlap fermions have been reported in this 
conference~\cite{Aoki:2017paw}. This study seems to suggest that the $\chi_t$ does not vanish linearly as $m_q$ but may 
either go as $m_q^2$ or rather abruptly vanishes for quark masses smaller than a critical mass $\lesssim 10$ MeV on a lattice 
of volume $~(2.4 ~fm)^3$. When the volume is increased to $\sim (3.6~fm)^3 $,  the value of $\chi_t$ at $m_q\sim10$ MeV 
increases to a non-zero value, whereas for even smaller masses it seems to be consistent with zero with larger errors. 
The gluonic definition of $\chi_t$, however gives a non-zero value even for masses $m_q<5$ MeV. It would be interesting 
to check if this difference in values of $\chi_t$ measured using gluonic and fermion methods as a function of quark mass 
is resolved as one goes to the infinite volume and continuum limits.

Calculating bulk thermodynamic quantities of QCD on the lattice has interesting developments in past couple of years, both in terms of 
new techniques and results. There are updates on the EoS using non-perturbatively $\mathcal O(a)$ improved Wilson fermions 
where gradient flow is employed to fix the renormalization of the energy momentum tensor~\cite{Suzuki:2013gza} on the 
lattice~\cite{Taniguchi:2016ofw}. 
Since the continuum limit is not yet achieved, the idea is to look at the plateau of relevant quantities as a function of $a^2/t$ 
where $a$ is the lattice spacing and $t$ being the flow time~\cite{Taniguchi:2016ofw}. The systematics of taking $t\rightarrow 0$ 
before continuum limit is studied in numerically inexpensive $SU(2)$ and $SU(3)$ gauge theories~\cite{Kitazawa:2017qab} and 
reported in this conference~\cite{gradlat}. At present results for entropy density and subtracted chiral condensates on lattices of 
size $a\sim 0.09$ fm, are consistent with improved versions of staggered fermions whereas the interaction measure has still large errors. 
Update on the measurement of chiral susceptibility has also been reported~\cite{baba}. New applications of gradient flow to fix the 
renormalization of energy momentum tensor correlators in full QCD with $m_\pi/m_\rho\sim 0.6$ has been discussed~\cite{tanilatt}. 
The correlators of $T_{12}$ show a plateau-like behavior near large Euclidean times  
$\tau \sim N_\tau/2$ for different flow times at $T=232$ MeV whereas for diagonal $T_{ii}$ correlators the plateau is  
quite noisy. Using model ansatz for spectral functions, the shear and bulk viscosities have been measured, the latter 
with larger errors. At present the results using the HTL ansatz cannot be differentiated from the Breit-Wigner ansatz for 
$T>200$ MeV and further spectral reconstructions are being studied~\cite{tanilatt}. A new result of the jet quenching parameter 
for $SU(3)$ gauge theory was reported~\cite{jet}. Precise measurements of these real-time coefficients~\cite{Pasztor:2018yae} 
will ultimately tell us how perturbative is the QCD medium beyond $T_c$.  This is an evolving area, where lattice techniques 
need further development and has a promising potential. For recent updates on measurements of other real-time quantities like 
photon and di-lepton rates see Ref.~\cite{Kaczmarek:2017hfx} and a talk in this conference~\cite{ding}.

The EoS of QCD has now been measured with HISQ fermions for temperatures upto $2$ GeV by carefully performing continuum 
extrapolation~\cite{peter}, results of which were discussed in this conference (left panel of Fig. \ref{fig:eos}). The results 
are consistent with expectations from 3-loop HTL perturbation theory (without the static magnetic contribution). In fact measurements of the screening correlators at finite temperature for mesonic excitations in QCD~\cite{scr} reveal that though in vector and axial-vector
channels, the convergence to their perturbative estimates is quick, the scalar-pseudo-scalar excitations have a very slow convergence 
towards the perturbative value. Larger symmetries $SU(2N_f)$ of fermion charge seem to be visible through the degeneracy of 
screening correlators of vector $V_x$ and tensor $T_t$ excitations as reported in~\cite{rohr} for $T>2~T_c$, when $U_A(1)$ is approximately 
restored.  Near the perturbative regime at $T\sim ~5~T_c$, these symmetries are again observed to be broken explicitly. 
All these studies hint to the fact that QCD medium is still non-perturbative beyond $T_c$ and the elementary excitations of the plasma 
have far more intricate structures than just free quark and gluon-like quasi-particles. The production of strange degrees of freedom 
is one of the other proposed signatures of a non-perturbative quark-gluon plasma. The FASTSUM collaboration have reported~\cite{fastsum} 
on the parity restoration in different strange baryon channels near $T_c$. Though the $S=1$ baryon parity partners becomes degenerate 
like the non-strange baryons immediately near $T_c$, for higher strangeness sectors the parity restoration seem to occur much slowly, 
at $T>T_c$.

\section{Towards Understanding the Columbia plot}
\begin{figure}[t!]
\begin{center}
\includegraphics[scale=0.2]{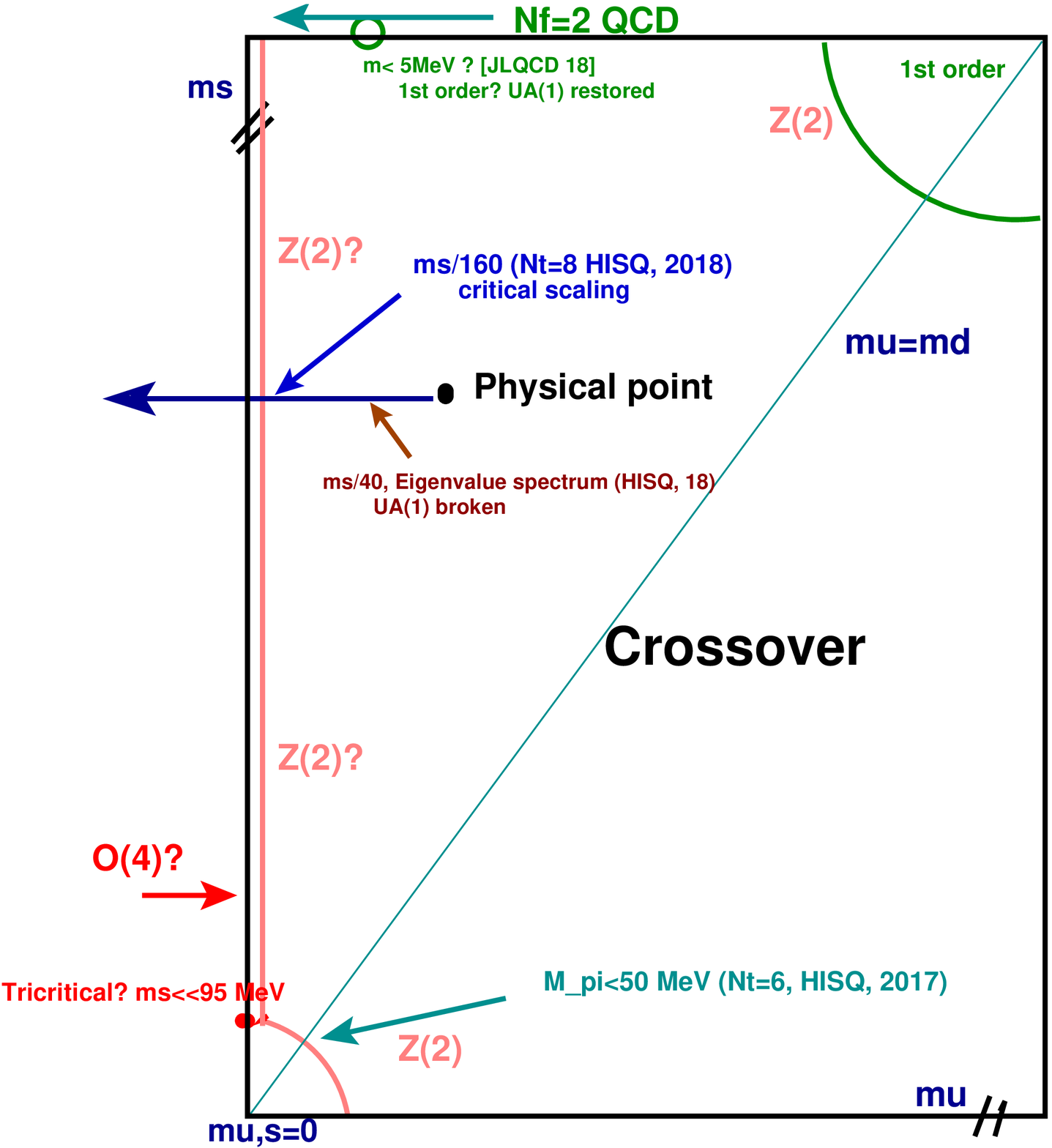}
\includegraphics[scale=0.55]{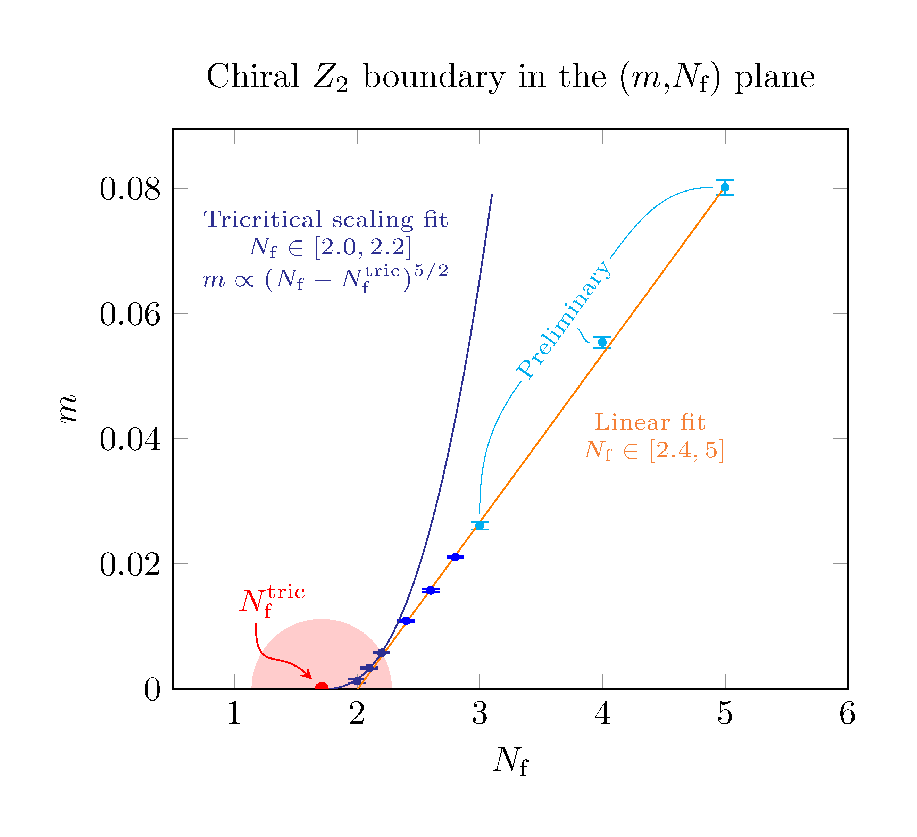}
\caption{The current status of Columbia plot from lattice studies (left panel). Right panel shows the tricritical 
scaling fit of quark mass as a function of $N_f$ from Ref.~\cite{Cuteri:2017gci}. }
\label{fig:cplot0}
\end{center}
\end{figure}

A deeper understanding of the phase diagram of QCD is obtained when one looks at a more fundamental problem: 
what is the fate of 'chiral' phase transition when the masses of quark flavors are varied. In left panel of 
Fig.~\ref{fig:cplot0} the current status of the famous Columbia plot is summarized. QCD with physical quark 
masses lie in the crossover region extended for a range of $m_u,m_s$. The upper right corner of the plot is 
much better understood since for quark masses infinitely large, it corresponds to $SU(3)$ gauge theory which 
has a first order transition. This first order region is separated from the crossover region by a $Z(2)$ second 
order line. The lower left corner is comparatively much less understood. From model QFTs with same symmetries as 
$N_f=3$ QCD, it is expected that a first order region exist which should again be separated from the crossover 
region by a second order $Z(2)$ line. In fact $Z(2)$ scaling studies of chiral susceptibilities along the diagonal 
with $m_s=m_{u,d}$ on $N_\tau=6$ lattices with HISQ fermions constrain the $Z(2)$ line to exist for pion masses 
$m_\pi<50$ MeV~\cite{Bazavov:2017xul}. With clover improved Wilson fermions the corresponding critical pion mass is $m_\pi<170$ 
MeV but for rather coarse lattices at present~\cite{Jin:2017jjp}. However for both staggered as well as Wilson fermions, 
the first order region tends to shrink when the lattice spacings are made finer. In a very insightful report~\cite{deForcrand:2017cgb} 
it has been motivated that as one approaches the continuum limit, the first order region for $N_f=3$ or even $N_f=4$ 
will shrink even further.  For more updates on the status of critical $m_\pi$ for $N_f=4$ QCD with Wilson fermions, see~\cite{ohno}. 
The other question that naturally arises in this context, is whether this first order region end at a tricritical point for 
$m_{u,d}=0$ and a finite $m_s$ or continues all the way to the $m_s\rightarrow \infty$ axis. Which of these two scenarios survive 
in the continuum limit may ultimately be related to the fate of $U_A(1)$, which is not yet conclusively known. Already with coarser 
lattices, the first order region seems to be quite tiny in the lower left corner of the plot. If indeed this first order region 
survive as a tiny strip parallel to the $m_s$ axis and continue to $m_s\rightarrow \infty$ i.e. $N_f=2$ axis, then 
it is expected that the corresponding $m_{u,d}$ is much smaller than physical quark masses. The arrows on the plot in 
Fig.~\ref{fig:cplot0} indicate the directions of some of the current lattice studies in this regard. Summarizing them, 

\begin{figure}[]
\begin{center}
\includegraphics[scale=0.25]{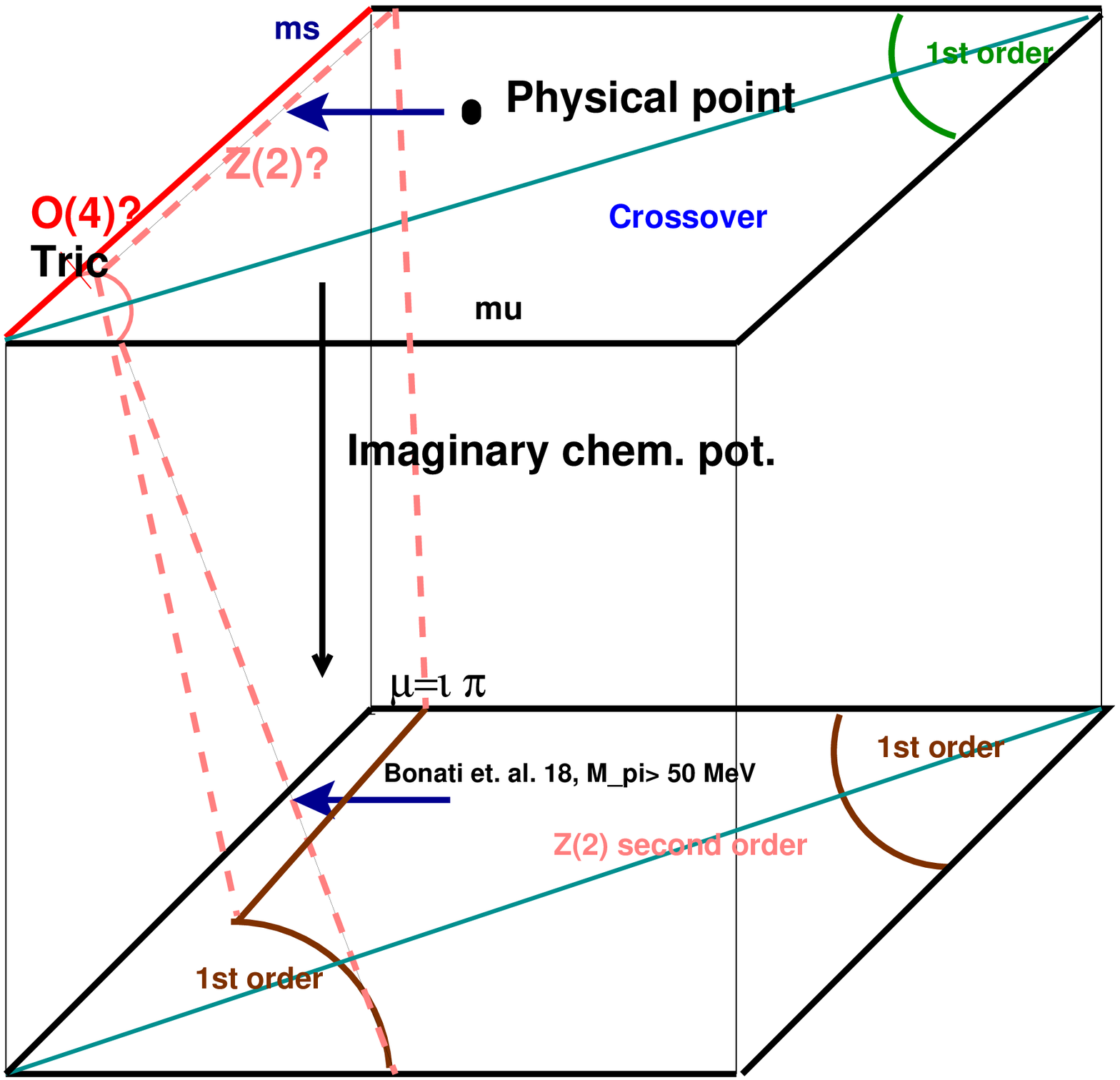}
\includegraphics[scale=0.5]{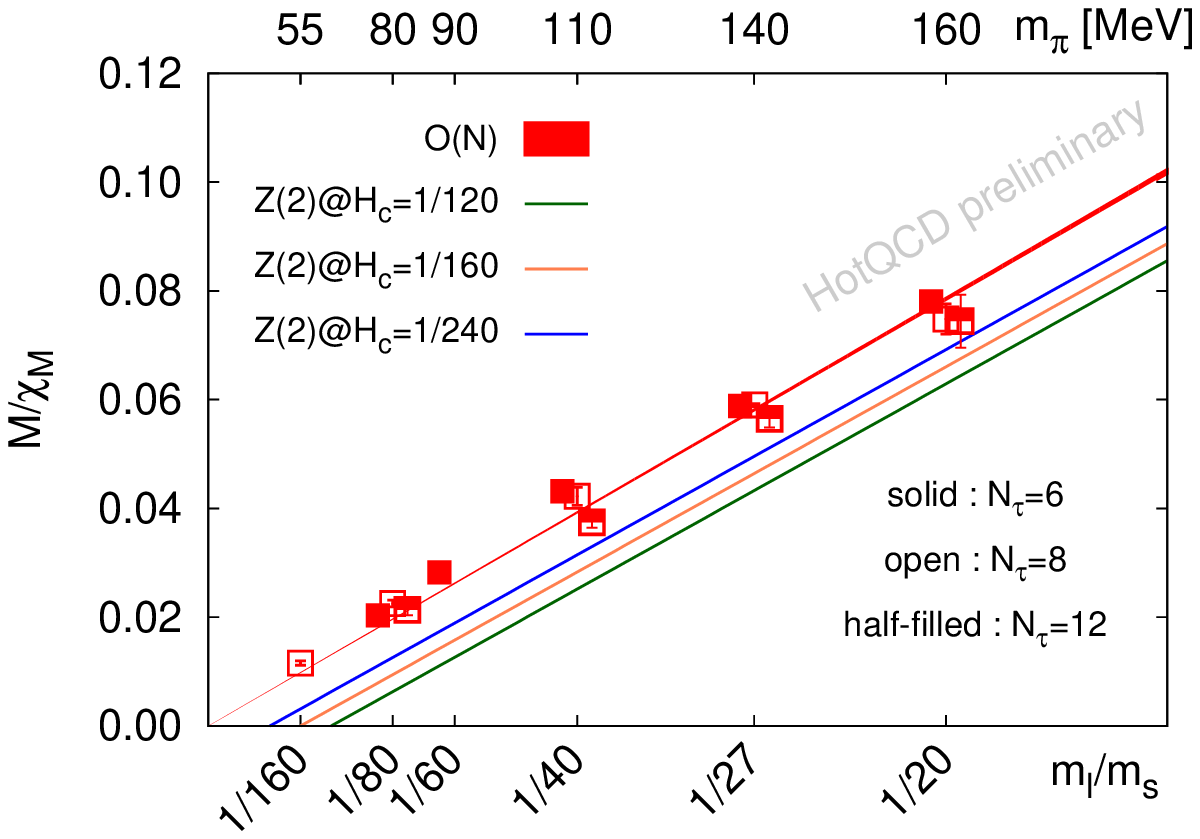}
\caption{Current status of the Columbia plot with an additional imaginary $\mu_B$ axis added (left panel). 
Right panel shows the scaling of the chiral condensate at $\mu=0$ as a function of light quark mass from 
Ref.~\cite{Ding:2018auz}. }
\label{fig:cplotmu}
\end{center}
\end{figure}

\begin{itemize}
 \item The green arrow shows the JLQCD approach for $N_f=2$ QCD, explained in the previous section~\cite{suzukilatt}. 
With the current lattice volume $(2.4~fm)^3$ and spacing $a^{-1}=2.6$ GeV, the results seem to suggest that $U_A(1)$ 
is restored for $m_{u,d}\lesssim 5$ MeV and could well be in the first order region. However the expectations from 
$N_f=3$ QCD seems to suggest that in the continuum, the first order region, if it survives and continues 
from the lower left corner all the way to the $N_f=2$ axis will very narrow characterized by $m_{u,d}<<5$ MeV. 
It will be important to reconcile both these results in the continuum limit. A related work discussed in this 
conference was to extract $T_c$ from a reweighted spectral density of QCD and thus obtain the order of transition 
in $m_{u,d}\rightarrow0$ limit~\cite{Endrodi:2018xto}.
\item The blue line on the Columbia plot shows the other approach by the HotQCD collaboration~\cite{Ding:2018auz}, 
where $m_s$ is fixed to its physical value and  $m_{u,d}$ successively reduced to check whether one approaches the 
$Z(2)$ line to the left or goes over to a $O(4)$ second order line. New results on chiral susceptibility $\chi_M$ 
for $N_\tau=8,12$ lattices with HISQ fermions discussed in the conference, suggest that the peak of $\chi_M$ 
decreases with volume ruling out first order phase transition for $M_\pi>80$ MeV. Scaling studies of the chiral 
condensate normalized by $\chi_M$ seems to rule out $Z(2)$ scaling for $M_\pi>55$ MeV, see right panel of 
Fig.~\ref{fig:cplotmu}. 
\item New studies on the eigenvalue distribution of $2+1$ flavor QCD with HISQ fermions discussed 
earlier~\cite{lukaslat} also follows along this blue arrow. It has a different motivation, 
to find out if $U_A(1)$ remains broken as light quark mass is successively reduced from its 
physical value. If indeed $U_A(1)$ is broken, the $Z(2)$ line will not exist and one will 
directly hit the $O(4)$ line when moving towards $m_{u,d}\rightarrow 0$. The eigenvalue 
densities as observed for 3 pion masses $M_\pi\sim 160, 140, 110$ MeV seem to support 
this latter scenario.
\end{itemize}

A more general approach has been discussed in this conference: to vary the $N_f$ as a continuous 
parameter and study the fate of the chiral phase transition~\cite{Cuteri:2017gci}. The idea is to 
start with $N_f=3$ QCD with finite quark masses in the first order region and zoom in to the tricritical 
scaling regime to extract $N_f^{tric}$ such that $m_q\sim (N_f-N_f^{tric})^{5/2}$. For $N_\tau=4$ 
lattices the $N_f^{tric}<2$, which seems to suggest a first order transition for $N_f=2$ (see right panel 
of Fig.~\ref{fig:cplot0}). These results are being further verified in the continuum limit. 

\subsection{Adding a new axis to the Columbia plot}
\begin{figure}[]
\begin{center}
\includegraphics[scale=0.35]{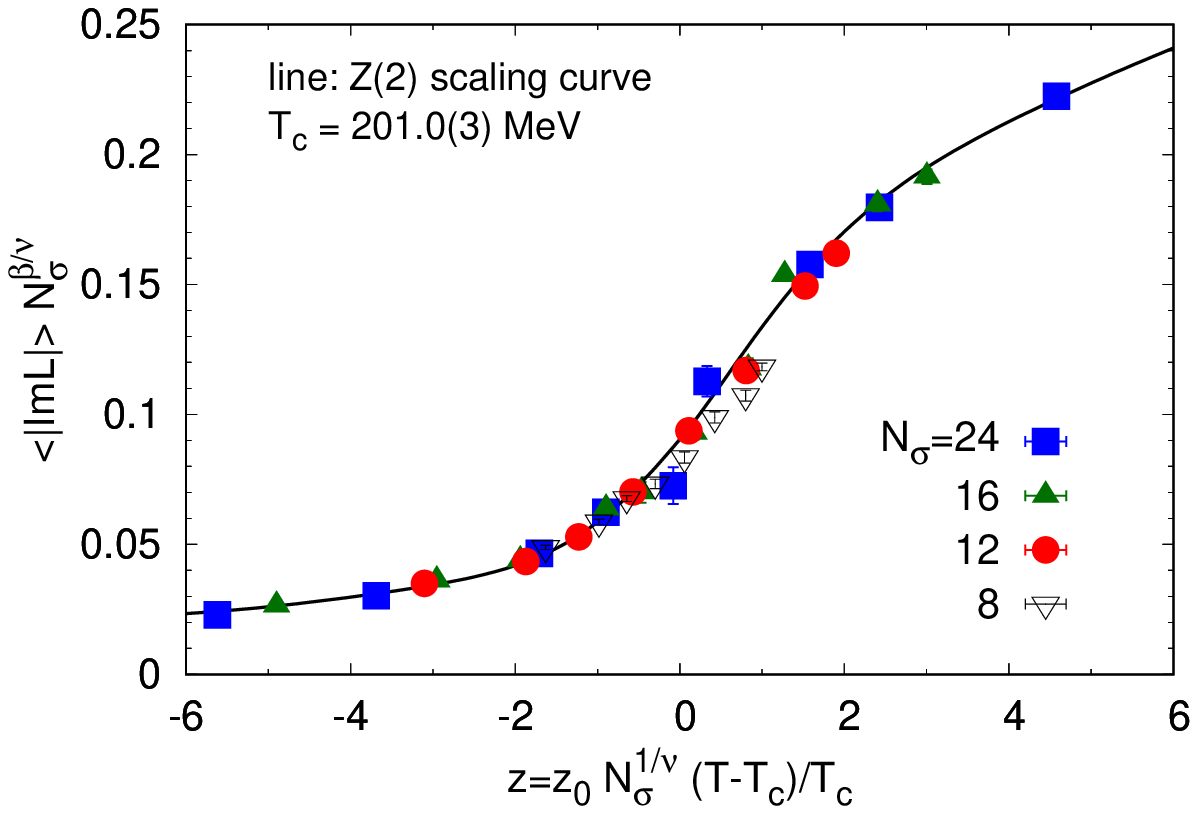}
\includegraphics[scale=0.33]{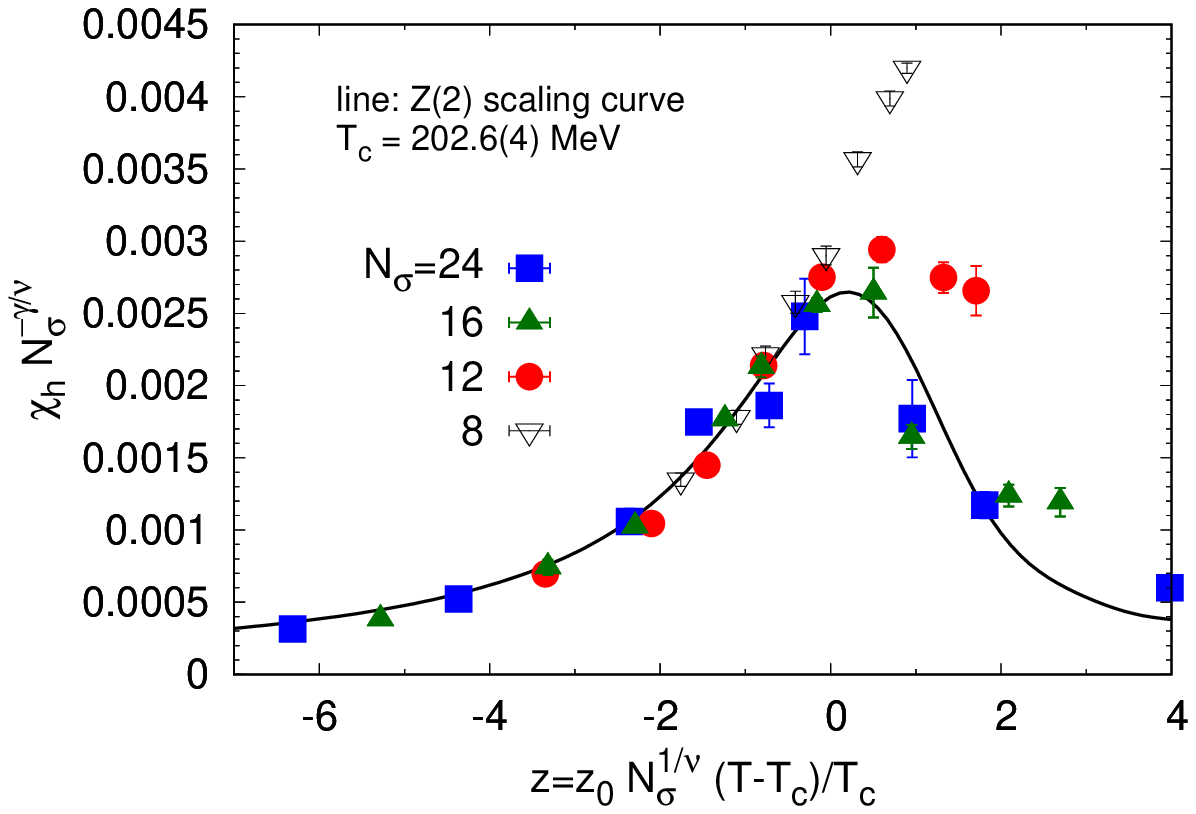}
\includegraphics[scale=0.2]{tc_fit.eps}
\caption{The left and the central panel shows the scaling of the imaginary part of Polyakov loop and its susceptibility respectively 
from~\cite{jishnu}. The right panel shows the agreement between $T_{RW}$ and the chiral transition temperature from~\cite{Bonati:2018fvg}.}
\label{fig:scimu}
\end{center}
\end{figure}

It was suggested in Ref.~\cite{deForcrand:2010he} that adding a third axis in form of an imaginary quark chemical 
potential $i\mu_q$ can further impose constraints on the 2D conventional Columbia plot. The QCD partition function 
in presence of $i\mu_q$ is free from 'sign-problem' and has symmetries $Z(\frac{\mu_q}{T})= 
Z(-\frac{\mu_q}{T})$ and $Z(\frac{\mu_q}{T})= Z(\frac{\mu_q}{T}+\frac{2n i}{3} \pi)$ for $n \in \mathcal Z$. 
The center symmetry is thus again a good symmetry even in presence of finite quark masses. The phase of the 
Polyakov loop is an observable in this case, which will identify the different $Z(3)$ sectors as $i\mu_q$ is 
varied. 
For the Roberge-Weiss (RW) points characterized by $\mu_q=(2n+1)\frac{i\pi T}{3}$~\cite{Roberge:1986mm}, there 
is a transition between adjacent center sectors, which is first order for high temperatures and a smooth crossover 
for lower temperatures. From continuity, the first order lines should end in a second order RW end-point. The interesting 
question is how the deconfinement and chiral transitions at $\mu_q=0$ connect to the RW point. For heavy quark masses
the first-order deconfinement lines, from reflection symmetry of the partition function, is expected to meet at the RW 
point, which will be a triple point.  For intermediate values of $m_q$, the crossover curve at $\mu_q=0$ may meet at the 
RW end-point, expected to be in $Z(2)$ universality class. The chiral limit is however more interesting. 
Numerical simulations, initially on $N_\tau=4$ lattices with staggered fermions have shown a first order 
RW transition for both $N_f=2,3$~\cite{DElia:2009bzj,deForcrand:2010he}, likely to survive in the chiral 
limit~\cite{Bonati:2014kpa}, confirmed later in studies with Wilson fermions~\cite{Cuteri:2015qkq}. This scenario is 
summarized in the modified Columbia plot at $\mu_q=\mu_B/3=i\pi T/3$, shown in left panel of Fig.~\ref{fig:cplotmu}. 
It is expected that the $Z(2)$ second order transition at intermediate masses is separated from the first order 
regions for both $N_f=2$ and $N_f=3$ by tricritical points. Now, what are its consequences for the $N_f=2$ chiral 
transition at $\mu_q=0$? If the $N_f=2$ chiral transition at $i\mu_q=0$ is, \\
a)~second order, then the first order RW transition will end in a tricritical point for $\mu_q^2<0$.\\
b)~first order, then the first order RW transition would 
 end in a tricritical point at $\mu_q^2>0$.\\
The first lattice study along this line~\cite{Bonati:2014kpa} was performed with staggered fermions on $N_\tau=4$ 
for different lattice volumes with $N_s=8,12,16$. The $Z(2)$ second order line was estimated for finite quark 
masses and for different values of $i\mu_q$ from Binder cumulants. Subsequently the $\mu_q^{tric}$ was estimated 
by looking at the tricritical scaling for $m_{u,d}$ in the chiral limit. The tricritical point was found at 
$\mu_q^2=0.85(5) T^2$ which seemed to suggest that $N_f=2$ chiral transition at $\mu_q=0$ is first order atleast 
on coarser lattices~\cite{Bonati:2014kpa}. 
Subsequently improved versions of staggered fermions are been used to reduce lattice cut-off effects, which play a 
decisive role in this study. The most recent high statistics studies are being performed for $2+1$ flavor QCD by keeping 
the $m_s$ fixed to its physical value and reducing the $m_{u,d}$ at $\mu_q=i\pi T/3$ along the blue line shown on 
the lower RW plane of Fig. \ref{fig:cplotmu}. To summarize these results:

\begin{itemize}
 \item Studies with stout-smeared staggered fermions have been performed for several lattice spacings $N_\tau=4-10$ 
 with  current state of the art being $a=0.1$ fm~\cite{Bonati:2018fvg}. The light quark masses have been varied such 
 that the lowest  pseudo-Goldstone pion mass achieved in the numerical studies is $50$ MeV. The largest volume is 
 $N_\sigma=32$ such that $M_\pi L >1$. From scaling studies of the Polyakov loop susceptibility, a first order RW 
 transition is not observed for $M_\pi\geq 50$ MeV, which in the continuum limit would imply that the first order 
 region, if it continues to the $\mu_q=0$ plane would be a very narrow strip parallel to the $m_s$ axis. The other 
 question addressed in this study is how close are the RW and the chiral transitions. As evident from right panel 
 of Fig. \ref{fig:scimu}, the chiral and RW transition seem to follow each other as one reduces the $m_{u,d}$. 
 The scaling studies of the subtracted chiral condensate near the RW point at present cannot distinguish between 
 $O(2)$ universality scenario for $N_f=2$ and the $Z(2)$ universality expected at the RW transition~\cite{Bonati:2018fvg}.

\item The RW transition is related to the restoration of $Z(2)$ symmetry. Under $Z(2)$ transformation, the real part of 
Polyakov loop does not change sign whereas its imaginary part changes sign. Hence the expectation value $\langle |Im L|\rangle$ 
is a good order parameter and will show $Z(2)$ scaling. Scaling studies performed with HISQ fermion discretization for 
$N_\tau=4$ and $N_\sigma=8-24$ around the chiral crossover transition temperature $T_c\sim 200$ MeV for $M_\pi=135-90$ MeV 
has been reported in this conference~\cite{jishnu}. As evident from left and central panels of Fig. \ref{fig:scimu} a 
beautiful agreement with second order $Z(2)$ scaling is observed both for the order parameter and its susceptibility 
again confirming the previous independent finding~\cite{Bonati:2018fvg}. 
 \end{itemize}

The results using different improved versions of staggered fermions are converging to an agreement with no indication 
for a first order transition in the vicinity of RW fixed points for $M_\pi\gtrsim50$ MeV. However as argued in~\cite{Bonati:2018fvg}, 
the other pion states in these studies are still quite heavy so it is important to revisit these studies on more finer lattices 
or with other fermion discretizations.

\section{Status of QCD Phase diagram at finite density}
Simulating QCD at finite density on the lattice is one of the most challenging problems in theoretical physics. 
New computational techniques and algorithms have been discussed in this conference in order to ultimately simulate 
dense and (cold) quark matter and understand the yet unexplored regions of the phase diagram. For relevant references 
and phenomenological applications for QCD at finite density, see the plenary talk by C. Ratti in this conference~\cite{ratti}.

If indeed a first order transition occur in cold and dense QCD following clues from Nambu-Jona-Lasinio model, 
it should end in a critical end-point since we now know for sure that there is a crossover transition at 
$\mu_B=0$. Lattice is essential to establish if a critical end-point exist and to draw 
lines separating the phases of dense QCD matter. In this section, I will rather discuss how existing 
lattice techniques are allowing us to draw the chiral crossover line at small $\mu_B$ 
and what promise it holds to reach all the way to the critical end-point. Out of the many methods developed 
over the years to circumvent the sign problem, two of them have now been adapted for simulations at large 
volumes and towards the continuum. One of them is to simulate QCD at imaginary $\mu_B<\mu_B^{RW}$, calculate thermodynamic 
quantities like baryon number density and extrapolate to the real $\mu_B$ plane~\cite{deForcrand:2002hgr,DElia:2002tig}. 
The other method is to calculate the partition function at $\mu_B=0$ and expand it as a Taylor series in $\mu_B$~\cite{Allton:2002zi}. 
If indeed singularities like a critical end-point exist in the $T-\mu_B$ plane, then its location will determine the radius of 
convergence of the series~\cite{Gavai:2004sd}. 

\begin{figure}[]
\begin{center}
\includegraphics[scale=0.4]{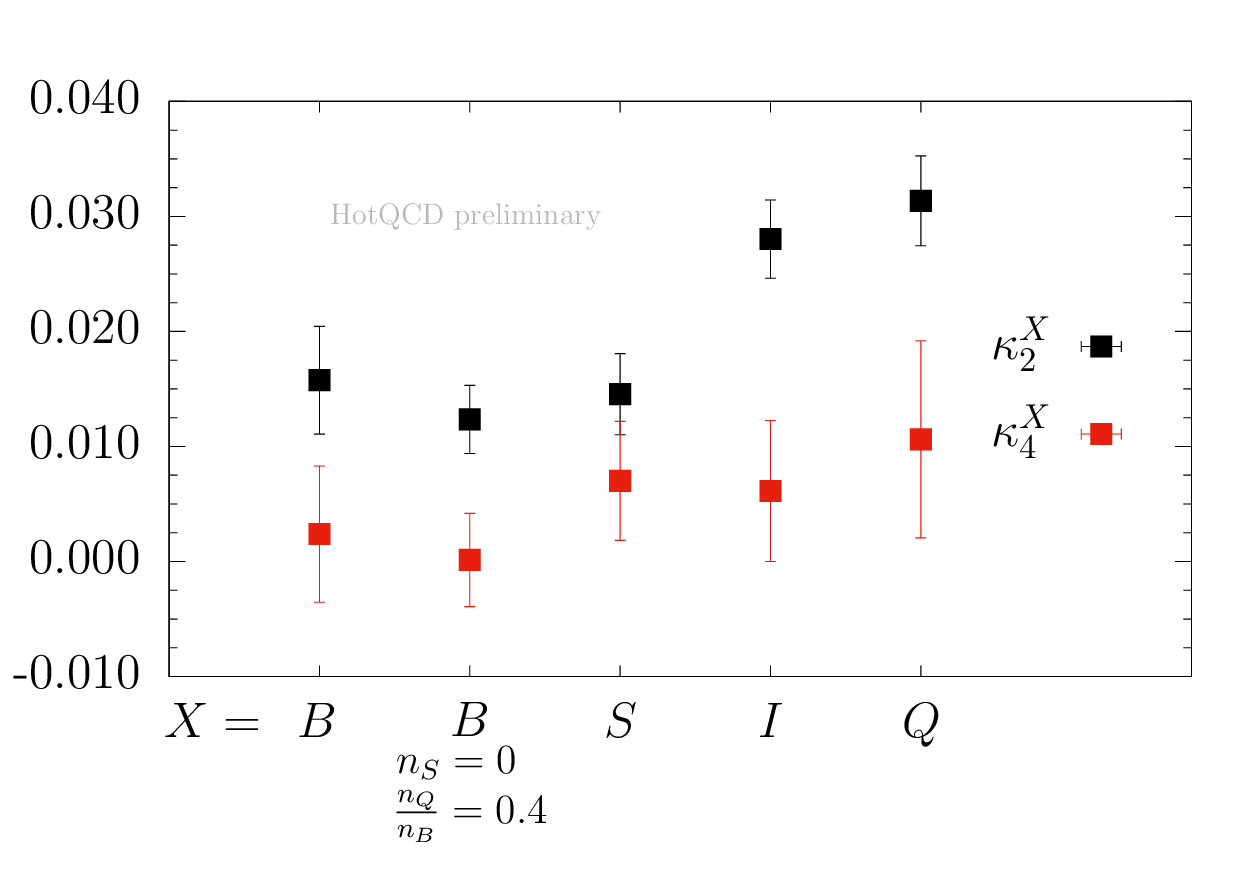}
\includegraphics[scale=0.2]{k_lqcd_new_new.eps}
\caption{The curvature for the pseudo-critical line calculated using HISQ fermions from Ref.~\cite{plattice} (left panel). 
The right panel shows the curvature estimates from different lattice groups from Ref.~\cite{DElia:2018fjp}.}
\label{fig:curv}
\end{center}
\end{figure}

\begin{figure}[]
\begin{center}
\includegraphics[scale=0.5]{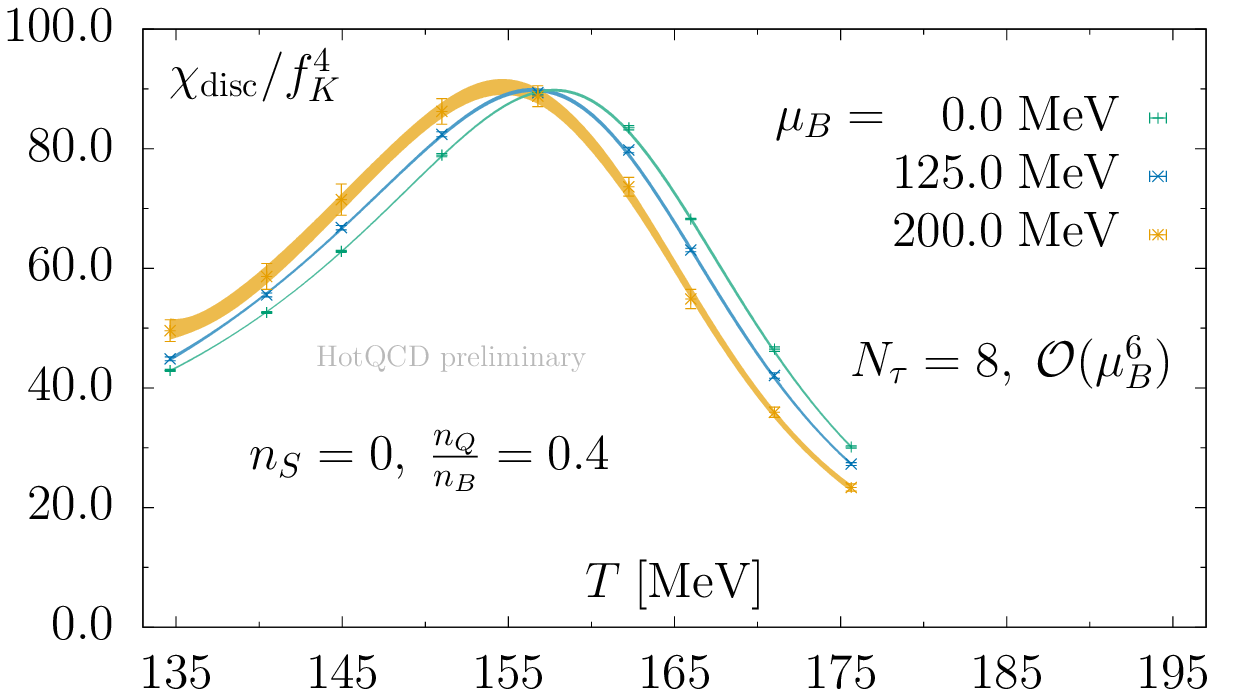}
\includegraphics[scale=0.5]{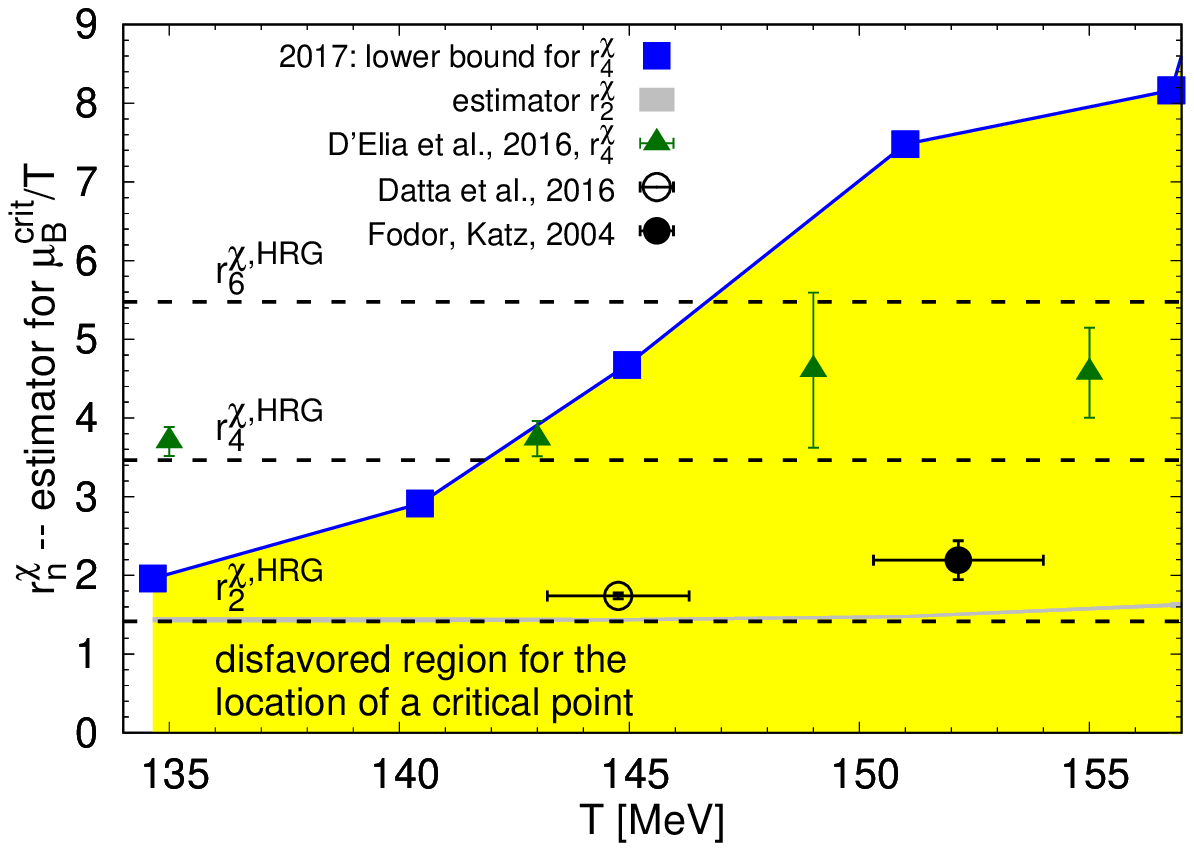}
\caption{Disconnected part of chiral susceptibility in QCD calculated upto $\mathcal O(\mu_B^6)$ using HISQ fermions on 
$32^3\times8$ lattice from ~\cite{Steinbrecher:2018phh,plattice} (left panel). A summary of the radius of convergence 
estimates from different lattice groups is shown in right panel from ~\cite{hotdense,Bazavov:2017dus}.}
\label{fig:Texp}
\end{center}
\end{figure}

Both methods have been used to calculate the curvature $\kappa_2$ and higher
derivatives $\kappa_4$ of the chiral crossover line at small $\mu_B$, defined through
$\frac{T_c(\mu_B)}{T_c(0)}=1-\kappa_2 \frac{\mu_B^2}{T_c(0)^2}-\kappa_4 \frac{\mu_B^4}{T_c(0)^4}$.
The results using Taylor expansion of chiral condensate with HISQ fermions for different $\mu_X$ 
where $X$ represents quantum numbers like baryon no., strangeness etc. were discussed in this 
conference~\cite{plattice} and summarized in the left panel of Fig. \ref{fig:curv}. 
The status of all recent lattice studies is summarized succinctly in the right panel of 
Fig. \ref{fig:curv} from Quark Matter 18 review by M. D'Elia~\cite{DElia:2018fjp}. 
For quite a few years, there was an apparent disagreement between the values of $\kappa_2^B$ obtained using 
Taylor expansion and imaginary $\mu$ methods. A careful continuum extrapolation was 
the key to this resolution~\cite{Bonati:2018nut}; the new continuum results with  HISQ fermions uses 
$N_\tau=6,8,12,16$~\cite{Steinbrecher:2018phh} data and with stout-smeared staggered fermions~\cite{Bonati:2018nut}
uses $N_\tau=6,8,10$ results. The values of $\kappa_2^B\sim 0.01$ and $\kappa_4^B\sim0$ suggest that the pseudo-critical 
line is almost flat for small $\mu_B$ and bends inwards very gradually towards larger $\mu_B$. Observables like the 
chiral disconnected susceptibility shown in left panel of Fig. \ref{fig:Texp} from Ref.~\cite{plattice,Steinbrecher:2018phh} 
also show a very mild dependence on $\mu_B$ for $\mu_B<250$ MeV.

New results for higher order fluctuations of conserved quantum numbers: baryon, charge, strangeness (B,Q,S) calculated using 
both these methods are available. For the imaginary $\mu$ method the latest high-statistics results are available 
from two different groups, both in $2+1$ QCD with stout-smeared staggered quarks. In one of these studies~\cite{DElia:2016jqh},
all possible diagonal and off-diagonal second order susceptibilities in the $(i\mu_B,i\mu_Q,i\mu_S)$ plane were calculated 
on $32^3\times8$ lattice for temperatures between 135-350 MeV. Approximating these second order correlations 
and fluctuations by a polynomial of $\mathcal O(\mu_B^i\mu_S^j\mu_Q^k)~,~i+j+k\leq8$ and extrapolating to the 
real plane, all higher order susceptibilities upto 8th order have been calculated. The other group uses a 
finer $48^3\times 12$ lattice to calculate all possible correlation and fluctuations of B, Q, S for temperatures 
between 135-220 MeV at 8 different imaginary $\mu$ values~\cite{Borsanyi:2018grb}. These were fitted to a polynomial of 
$\mathcal O(\mu_B^i\mu_S^j\mu_Q^k)~,~i+j+k\leq10$, where the eighth and tenth order data were put in as priors. Using this 
fitting procedure, higher order susceptibilities upto 8th order were reported. Since there is a discontinuity of the 
imaginary baryon number density at the first RW point for $T\geq T_{RW}$, this naturally limits the number of imaginary $\mu$'s where 
the simulations can be performed and hence the extrapolation to real $\mu$.  Given that range of simulations are 
limited to $\mu_B/T\in [0,i\pi)$, this method works better for $T<T_c$ but the systematic errors start dominating 
for $T>T_c$. In the Taylor expansion method, the pressure is expanded as a series in $\mu_B/T$ where the expansion 
coefficients are $\mu_B$-derivatives of pressure i.e. the higher order susceptibilities, calculated at $\mu_B=0$. These 
quantities involve derivatives of Dirac operator and each derivative is associated with an inverse of the 
Dirac matrix. The higher order fluctuations thus contains many such terms with alternating signs for subtle 
cancellations of the divergences to give a finite result. For $\chi_6^B$ or higher, the divergences may not 
exist which allows using a different technique~\cite{Gavai:2014lia} to compute them, which is computationally 
much cheaper compared to the conventional method~\cite{Hasenfratz:1983ba}. A new numerical implementation of 
this technique which may allow to calculate even higher order fluctuations was discussed~\cite{jaeger}.  The 
current state-of-the-art results using Taylor expansion are correlations and fluctuations upto sixth order using 
Highly Improved Staggered quarks~\cite{Bazavov:2017dus} and upto eighth order using unimproved staggered 
fermions~\cite{Datta:2016ukp}. Even with susceptibilities upto $\mathcal O(\mu_B^6)$, the QCD EoS show very good 
convergence for $\mu_B/T\lesssim 2.5$, the continuum estimates of which can be found in~\cite{Bazavov:2017dus}.

For locating the critical end-point, the radius of convergence (RC) of the Taylor series of 
pressure or the baryon number fluctuation, $\chi_2^B$ has to be estimated. From the definition of RC
 $r_{2n} \equiv \sqrt{2n(2n-1)\left|\frac{\chi_{2n}^B}{\chi_{2n+2}^B}\right| }$ it is not known a-priori how 
 large the order $n$ should be chosen in order to reliably extract this quantity on the lattice.
The current estimates of the radius of convergence is summarized in right panel of Fig. \ref{fig:Texp}. 
Most of them except the reweighting data are from $N_\tau=8$ lattices. The $r_4$ already deviates from the 
Hadron Resonance gas model (HRG) estimates by $\sim 30\%$ for $T\sim145$ MeV~\cite{Bazavov:2017dus}. There 
is a substantial difference between the $r_2$ and $r_4$ estimates so one needs atleast $r_6$ to get a reliable 
prediction for the RC. The yellow 'exclusion' region comes from the upper error bar of the $\chi_6^B$ 
measured using the HISQ fermions whose central values are given by the blue points~\cite{Bazavov:2017dus,hotdense}. 
Results using stout-smeared staggered quarks~\cite{DElia:2016jqh} also favor a larger $\mu_B^{CEP}/T$ than using 
the standard staggered quarks~\cite{Datta:2016ukp} shown by the black solid point. All these results should ultimately 
agree in the continuum limit. The $\mu_B^{CEP}/T$ using reweighting techniques from Ref.~\cite{Fodor:2004nz}  
favors a lower value; it will be interesting to confirm this in the thermodynamic limit. To summarize, the present 
lattice data for $\chi_n^B$ already  deviates from naive expectations from HRG model at $T>145$ MeV; the 
higher the order $n$, the more visible is the deviation. Moreover the present lattice data favor a small curvature 
of the chiral crossover line~\cite{Bonati:2018nut,plattice}. Furthermore the fact that $\kappa_4^B\sim0$ suggest if a CEP exist 
in the phase diagram then $T_{CEP}\lesssim T_c(\mu_B=0)$. For the case $T_{CEP}/T_c(0)\sim 0.95$, lattice data already suggests 
stronger departure from HRG results and can at present provide a suggestive lower bound, $\mu_B^{CEP}> 4 T$~\cite{hotdense}; the 
convergence to the actual value will depend on a more precise calculation of $\chi^B_8$. On the other hand if $\kappa_6^B$ 
or $\kappa_8^B$ have strong contribution to the curvature of the pseudo-critical line and $T_{CEP}/T_c(0)\leq 0.9$, the RC 
estimates would be more closer to HRG values, hence will show extremely slow convergence as a function of the order $n$.

\section{Outlook}
In this review, I hopefully could convince that the quest to understand the phase diagram of QCD has 
led to the many interesting theoretical and algorithmic developments in lattice gauge theory. Efforts 
to understand the chiral transition better, has led to a rich theoretical knowledge of QCD in the $m_{u,d}$-$m_s$ 
plane, additionally along the imaginary chemical potential as a new axis and now even as a function of $N_f$ as a 
continuous parameter. The role of $U_A(1)$ anomaly on the chiral phase transition is not yet fully understood but 
has led to many new insights on the microscopics of the QCD Dirac operator and its intimate connections to topology.  
The topological structures in QCD and their interactions is long suspected to drive chiral symmetry breaking 
and confinement at finite temperature and/or densities; new insights on which are coming from lattice studies. 
Moreover a strong motivation to quantify topological fluctuations at high temperatures have led to recent developments 
of interesting new algorithms that has even more wider applicability for lattice simulations near the continuum limit. 
The quest to go deeper along the $\mu_B$ axis of the phase diagram has triggered developments of algorithms and new techniques 
to circumvent the sign problem, with initial bounds available from lattice towards constraining the location of the critical 
end-point. The EoS characterizing different phases of QCD upto $\mu_B/T\lesssim 2.5 $ is now available; an 
increasing sophistication of lattice techniques is leading towards quantifying its dynamical properties.

\textbf{Acknowledgements}
I thank the organizers for the kind invitation.  Also I express my gratitude to Swagato Mukherjee who kindly agreed to fill in for my 
absence at the conference due to unavoidable circumstances, at a short notice and share his perspectives. I thank all the researchers 
working in finite temperature and density lattice QCD who have generously shared their inputs and results, specially to Massimo D'Elia, and also for the exciting talks in LATTICE 18; 
it indeed shows how vibrant this field of research is. 
Last but not the least, I am grateful to the members of HotQCD collaboration for a very enjoyable collaboration over these years and their 
helpful feedback.

\end{document}